\documentstyle[twoside,fleqn,espcrc2,epsf]{article}
\begin{document}

\title{Tau neutrinos from active galactic nuclei}

\author{Athar Husain\\
{\ }\\
  Departamento de F{\'\i}sica de Part{\'\i}culas,\\ Universidade de 
 Santiago de Compostela, E-15706 Santiago de
 Compostela, Espa{\~n}a}

\begin{abstract}

	We study the appearance of $\nu_{\tau}$ from active galactic 
 nuclei (AGN) through neutrino spin-flip. We consider two situations:
 i) Spin (flavour)-precession in a reasonable strength of magnetic 
 field in AGN, $B_{AGN}$ ii) Adiabatically resonant conversions caused 
 by an interplay of $B_{AGN}$ and the violation of equivalence 
 principle. Observational consequences for both situations 
 are discussed.

\end{abstract}

\maketitle

\section{INTRODUCTION}

	Neutrinos are currently understood as elementary particles 
having {\em spin} 1/2 ($\hbar$). Neutrinos have no colour charge, 
no electromagnetic charge, however neutrinos have weak charge 
 and consequently neutrinos interact only weakly. Light stable neutrinos 
come in three flavours, namely, $\nu_{e}$, $\nu_{\mu}$ and $\nu_{\tau}$
 alongwith corresponding antiparticles. Exact value of the mass of neutrino 
 is still unknown. All direct and indirect searches so far imply an
 upper bound on neutrino masses. The nonzero mass of the neutrino, however,
 on quite general grounds imply nonzero neutrino magnetic moment, 
 $\mu $ \cite{YU}. We intend to discuss here a phenomenological implication
 of $\mu$ in $B_{AGN}$.

	Some galaxies have quite bright centers. The luminosity of these 
galaxies typically reach $10^{44}-10^{48}$ erg/sec. In general, AGNs refers 
to these bright centers. It is hypothesized that the existence of a 
supermassive black hole $(M\, \sim (10^{6}-10^{10})\, M_{\odot})$ 
 [here $M_{\odot}$ is solar mass; $M_{\odot} \, \sim \, 2\cdot 10^{33}$ g] 
 may explain
the observed brightness as this supermassive blackhole captures the matter
 around it through accretion. This supermassive blackhole is hypothesized to
be formed by the collapse of a cluster of stars. During and after accretion, 
accelerated protons may collide with other protons and/or with the ambient
 photon field present in the vicinity of an AGN or/and in the associated 
jets (see latter in this Sect.) and produce unstable hadrons. These unstable
hadrons decay mainly into neutral and charged pions. The neutral pions 
decay into photons and thus may explain the observed brightness whereas the
charged pions mainly decay into electron and muon neutrinos and 
 corresponding antineutrinos. For an update review of various flux estimates 
 for $\nu_{e}$ and $\nu_{\mu}$ from AGNs (as well as from some other
 interesting astrophysical sources), see \cite{pro}. The same protons 
 (hypothesized to be present in the vicinity of an AGN) and photons 
  may give rise
to $\nu_{\tau}$ (and $\bar{\nu}_{\tau}$) in a small number, through, for
instance, $p+\gamma \, \rightarrow \, D^{+}_{S}+\Lambda^{0}+\bar{D}^{0}$.
 It is so because the production cross for $D^{+}_{S}$ is essentially two
orders of magnitude lower than, for instance, that of $\Delta^{+}$ (a main
source of $\nu_{e}$ and $\nu_{\mu}$) for the relevant center of mass energy
scale. The branching ratio of $D^{\pm}_{S}$ to decay eventually into
 $\nu_{\tau}$ ($\bar{\nu}_{\tau}$) is approximately an order of magnitude
 lower than for $\Delta^{+}$ to decay into $\nu_{e}$ and $\nu_{\mu}$. These
two suppression factors alongwith the relevant kinematic limits give
 approximately the ratio of fluxes of tau neutrinos and muon (anti) neutrinos
 as $\nu_{\tau}/\bar{\nu}_{\mu}\, {\buildrel < \over {_{\sim}}} \, 10^{-3}$
 \cite{vaz}.

Here, we discuss a possibility to enhance this ratio, that is, to obtain,
 $\nu_{\tau}/\bar{\nu}_{\mu}\, >\, 10^{-3}$, through neutrino spin-flip 
 induced by violation of equivalence principle (VEP) \cite{mug}.  
The VEP arises as different flavours of neutrinos may couple 
differently to 
gravity \cite{G,L}. This essentially results from the realization that 
flavour eigenstates of neutrinos may be the 
admixture of the gravity eigenstates of neutrinos with different 
gravitational couplings. 
	We study here mainly the spin-flip for Majorana type neutrinos 
 in the
vicinity of the cores of active galaxies which we hereafter refer to 
 as AGNs.
 Some AGNs give off a jet of matter that stream out from the nucleus 
 in a 
 transverse plane and produces hot spots when the jet  strikes the 
 surrounding 
 matter at its other end. For a discussion of neutrino spin-flip in 
 jets 
and hot spots, see \cite{DKD}.
	Previously, the neutrino spin-flip effects for AGN neutrinos 
 due 
to VEP 
are studied in \cite{R,W,SB}. The VEP is parameterized by a 
 dimensionless
parameter $\Delta f$. In \cite{R}, by demanding an adiabatic 
 conversion to 
occur, a lower bound on $\mu$ was obtained in terms of $\Delta f$. 
 In \cite{W}, neutrino spin-flip in
AGN due to gravitational effects (based on a certain choice 
 of the metric 
parameters) and due to the presence
of a large magnetic
field is studied, whereas in \cite{SB}, the effect of possible random 
fluctuation in  
 $B_{AGN}$ on neutrino precession is considered. There is no detailed 
 study of 
the implications of spin-flip 
effects for  neutrinos with small $\delta m^{2}$  originating
from AGN without gravitational effects and/or VEP.
 In this context, here we address two 
different aspects of spin-flip for  high energy neutrinos, viz, the  
spin (flavour)-precession
 with/without VEP in  $B_{AGN}$;  
  and the adiabatic/nonadiabatic conversion due to 
 an interplay of  $B_{AGN}$ and the VEP. We point out that,  
for latter  type of conversion effect,  
 a $\Delta f$ of the order of $10^{-39}-10^{-29}$ depending on 
$\delta m^{2}$ gives
 reasonably large conversion probabilities 
 for a suitable choice of parameters of the AGN model.

	The present study is particularly welcome as the new 
 ice/underwater 
\v{C}erenkov light  detectors namely AMANDA, Baikal, NESTOR and 
 ANTARES will
have  not only the energy but also the  angle and flavour 
 resolutions for high energy 
 neutrinos from AGNs \cite{C}. 
 These characteristics 
make these neutrino telescopes especially suitable for the study of 
  various high
energy neutrino conversions.

The plan of the paper is as follows. In Sect. 2, we briefly discuss 
the matter density and magnetic field 
profiles in AGN. In Sect. 3, we discuss the 
spin (flavour)-precession due to VEP and determine the value of 
 $\Delta f$ 
needed to have the precession  probability greater than 1/2. In 
 the same Sect., we consider in some
detail the resonant/nonresonant conversions induced by an interplay 
 of   
$B_{AGN}$, and the VEP alongwith corresponding observational 
 consequences and finally in Sect. 4, we summarize our results.

\section{THE MATTER DENSITY AND MAGNETIC FIELD IN AGN}

Neutrino spin-precession in the context of the Sun was discussed 
 in \cite{V}. 
It was pointed out that the  matter effects tend to suppress the 
precession effect. As shown below, for AGN,  matter effects
arising due to coherent forward scattering of neutrinos off the 
 background
are negligible (similar estimate for other astrophysical 
 systems 
like Sun and Collapsing Stars shows that the matter effects are 
 indeed 
nonnegligible in most part of these systems).
	The essential conditions needed for appreciable 
 spin-precession are:
 i) $\mu Br\, {\buildrel > \over {_{\sim}}} 1$, i.e., $B$ must be 
 large 
enough in the region $r$; ii) the smallness of the 
 usual matter effects, so that neutrino spin-precession is not 
 suppressed 
 (see below); 
and iii) there should be no reverse spin-precession of neutrinos 
 on their way
to earth. 
 As for the third essential condition the observed 
 intergalactic magnetic 
field for the nearby 
 galaxies is
estimated to be $\sim \, O(10^{-9}$G) at a scale of Mpc. Taking 
 a typical distance between the 
earth and the AGN as $\sim \, $100 Mpc, we note that the effect 
 induced by 
intergalactic/galactic magnetic field is quite small as the 
 galactic 
 magnetic field is $\sim \, O(10^{-6}$G), 
thus causing negligible reverse spin-precession.

According to \cite{S}, the matter density in the vicinity of AGN 
has the following profile: 
$\rho (x)\, =\, \rho_{0}f(x)$ where $\rho_{0} \simeq 
 1.4\cdot 10^{-13}$
g/cm$^{3}$ and $f(x)\, \simeq \, x^{-2.5}(1-0.1 x^{0.31})^{-1}$ 
as we take the AGN luminosity to be $10^{46}$ erg/sec with $x\,
\equiv r/R_{S}$, $R_{S}$ being the Schwarzchild radius of AGN: 
 $R_{S}\,
\simeq 3\cdot 10^{11}\left( \frac{M_{AGN}}{10^{8}M_{\odot}} \right)$ 
 m. We
 take the distance traversed by the neutrinos to be 
$10 < \, x\, < \, 100$ in the vicinity of AGN. These imply that
the width
of the matter traversed by  neutrinos in the vicinity of the AGN is 
$l_{AGN}\, \sim \, (10^{-2}-10^{-1})$ g/cm$^{2}$. In the presence of 
 matter, 
the effective width of
matter needed for appreciable spin-flip, on the other hand, is 
$l_{\mbox{o}}\, \equiv \sqrt{2}\pi m_{N}G^{-1}_{F} \, \sim \, 2 
\cdot 10^{9}$ g/cm$^{2}\, \gg l_{AGN}$, much larger than $l_{AGN}$. 
Hence, from now on, we ignore the matter effects.

We consider now the magnetic field in the vicinity of AGN with the 
following profile \cite{S}
\begin{equation}
 B_{AGN}(x)\, =\, B_{0}g(x),
\end{equation}
where $B_{0}\, \sim 5.5 \cdot 10^{4}$ G and 
$g(x)=x^{-1.75}(1-0.1x^{0.31})^{-0.5}$ for $10\, <\, x\,<\, 100$. 
 We will use this $B_{AGN}$ in our estimates.

\section{NEUTRINO SPIN-FLIP DUE TO VEP}

The evolution equation for the two 
neutrino state for vanishing gravity mixing angle and the vanishing
vacuum mixing
angle is governed by an effective hamiltonian whose diagonal elements 
may be written as \cite{Ak,A}
\begin{eqnarray}
  H_{\bar{\mu}} & = &   0, \nonumber \\
  H_{\tau}      & = &   \delta -V_{G}.
\end{eqnarray}
In Eq. (2), 
 $\delta \, =\, \delta m^{2}/{2E}$, where 
$\delta m^{2}\, =\, m^{2}(\nu_{\tau})-m^{2}(\bar{\nu}_{\mu})\,>\, 0$
 with $E$ being the neutrino energy. 
 $V_{G}$ is the effective potential felt by the neutrinos at a 
 distance $r$ 
from a gravitational source of mass $M$ due to VEP and is 
 given by \cite{G}
\begin{equation}
 V_{G}\, \equiv \, \Delta f\phi (r)E,
\end{equation}
where $\Delta f \, =\, (f_{3}-f_{2})(f_{3}+f_{2})^{-1}$ and 
$\phi (r)\, =\, G_{N}Mr^{-1}$ 
is the gravitational potential in the Keplerian approximation,
with $f_{2}G_{N}$ and $f_{3}G_{N} $being respectively the gravitational 
couplings for $\bar{\nu}_{\mu}$ and $\nu_{\tau}$, such that 
$f_{2}\neq f_{3}$.   
 The possibility of vanishing gravity and vacuum mixing angles in 
Eq. (2) allows us to identify the range of $\Delta f$ relevant for
 neutrino magnetic
 moment effects only.
We now propose to study  in some detail the various 
 possibilities arising from the 
relative comparison
between $\delta$ and  $V_{G}$  in Eq. (2).

{\noindent \bf Case 1.} $V_{G}\, =\, 0$. For 
 constant $B$, 
we obtain the following expression
for precession probability $P(\bar{\nu}_{\mu}\rightarrow \nu_{\tau})$ 
 from Eq. (2):
\begin{equation}
 P(\bar{\nu}_{\mu}\rightarrow \nu_{\tau}) = 
 \left[\frac{(2\mu B)^{2}}{(2\mu B)^{2}+X^{2}}\right]
 \sin^{2}\left(\frac{r\sqrt{D}}{2}\right),
\end{equation}
with $X\, =\, \delta$ and $D$ being the denominator in the prefactor of phase 
. We now discuss the relative comparison between 
 $2\mu B$ 
and $\delta $ and evaluate
$P$
for corresponding $\delta m^{2}$ range.

{\noindent a)} $\delta \ll 2\mu B$. Using $B$ given in Eq. (1) for 
$\mu \, \sim \, 10^{-12}\, \mu_{B}$
\cite{F}, we
obtain $\delta m^{2}\, \ll \, 5\cdot 10^{-4}$ eV$^{2}$ with 
$E\, \sim \, 1 $ PeV. We take here $\delta m^{2}\, \sim \, 5\cdot 10^{-6}$
 eV$^{2}$ as an example.  The expression (4) for $P$ then reduces to   
\begin{equation}
 P(\bar{\nu}_{\mu}\rightarrow \nu_{\tau})\, \simeq \, \sin^{2}(\mu B r). 
\end{equation}
The phase in $P$ can be of the order of unity if $\mu Br \, =\, 
\frac{\pi}{2}$ or if $\mu Br\, {\buildrel > \over {_{\sim}}} 1$ for a 
contant $B$.
Evidently, this $P$ is independent of $E$.  According to Eq. (1), the 
$B_{AGN}$ varies with distance so that to have maximal depth of precession, 
we need 
to 
integrate the strength of the magnetic field along the neutrino trajectory. 
Thus, we require 
\begin{equation}
 \int_{0}^{r} dr' B(r') \, {\buildrel > \over {_{\sim}}} \, \mu^{-1}.
\end{equation}
We note that Eq. (5) [alongwith Eq. (6)] give $P(\bar{\nu}_{\mu}\rightarrow 
 \nu_{\tau})\, > \, 1/2$  for the $B_{AGN}$ profile given by Eq. (1) 
with $\mu \, \sim \, 10^{-12}\, \mu_{B}$.  
Thus, an energy independent permutation (exchange) between $\bar{\nu}_{\mu}$
and $\nu_{\tau}$ may result with $P>\, 1/2$. 
 This energy {\em independent} permutation of energy spectra of 
 $\bar{\nu}_{\mu}$ and $\nu_{\tau}$ for
small $\delta m^2$  follows from the fact that Eq. (5) also gives 
$P(\nu_{\tau}\, \rightarrow \, \bar{ 
 \nu}_{\mu}$) as  we are considering a two neutrino state system.  
Let us further note that this small value of 
$\delta m^{2}$ ($\delta m^{2}\, \sim \, 5\cdot 10^{-6}$ eV$^{2}$) 
 is not only
interesting in the context of SNP \cite{AKH} but also SN \cite{A}.\\ 
{\noindent b)} $\delta \, \simeq \, 2\mu B$. Here $\delta m^{2}$ 
correspond to $5\cdot 10^{-4}$ eV$^{2}$. In
this case expression (4) for $P$ reduces to 
\begin{equation}
 P(\bar{\nu}_{\mu}\rightarrow \nu_{\tau})\, \simeq \, 1/2\, 
 \sin^{2}(\sqrt{2}\mu Br).
\end{equation}
Thus, for $\delta m^{2}\, \simeq \, 5\cdot 10^{-4}$ eV$^{2}$, energy 
dependent distortions may
result in {\em survived} and {\em precessed} neutrino energy spectra 
with $P\, {\buildrel < \over {_{\sim}}} \, 1/2$.\\
{\noindent c)} $\delta \gg \, 2\mu B$, that is, 
$\delta m^{2}\, \gg \, 5\cdot 10^{-4}$ eV$^{2}$. Energy
dependent distortions may result for large 
$\delta m^{2}$ with $P\, <\, 1/2$. For instance, consider 
$\delta m^{2} \, \sim \, 10^{-3}$ eV$^{2}$ relevant for 
 atmospheric 
neutrino problem \cite{YU}. The result of   
 vacuum oscillations with non vanishing vacuum mixing angle is   
a modification in the  
 $\bar{\nu}_{\mu}$ and $\bar{\nu}_{\tau}$ spectra through 
$\bar{\nu}_{\mu}\rightarrow
\bar{\nu}_{\tau}$ or a modification in $\nu_{\mu}$ and $\nu_{\tau}$ 
 spectra 
 through $\nu_{\mu}\rightarrow \nu_{\tau}$, whereas the 
 magnetic moment effects give rise to changes in $\nu_{\mu}$ 
 and $\bar{\nu}_{\tau}$ 
 spectra  or in $\bar{\nu}_{\mu}$ and $\nu_{\tau}$ spectra.
 This simply follows from the fact that a Majorana type neutrino 
 magnetic
 moment causes a precession between the relevant neutrino states of 
 opposite
 helicity as well as flavour in an external magnetic field.  
Thus, an empirical {\em distinction} 
 between high energy 
 $\nu_{\mu}$ and $\bar{\nu}_{\mu}$ spectra
 as well as between high energy 
$\nu_{\tau}$ and $\bar{\nu}_{\tau}$ spectra, if observed,
 will be an evidence for the magnetic moments effects here.   
The situation of $\bar{\nu}_{e}\, \rightarrow \, \nu_{\tau}$ 
may be realized by replacing $\bar{\nu}_{\mu}$ by $\bar{\nu}_{e}$ 
 in Eq.
(2) with the corresponding $\delta m^{2}$. A relevant remark is in 
order here: the deficit measured by superkamiokande in atmospheric 
 muon 
 neutrino flux may currently be explained either through 
 $\nu_{\mu}\rightarrow \nu_{\tau}$ or through $\nu_{\mu}\rightarrow 
 \nu_{s}$. For high energy neutrinos originating from AGNs, in the case 
 of $\nu_{\mu}\rightarrow \nu_{\tau}$, the ratio $\nu_{\tau}/\nu_{\mu}$ 
 is close to 1/2 as we take the $\delta m^{2}$ and the vacuum mixing angle 
suggested by recent superkamiokande data. Therefore, a ratio of 
 $\nu_{\tau}/\nu_{\mu}$ different from $\sim $1/2 
 correlated to the direction of 
 source may here provide an evidence for neutrino spin-flip in the AGN 
 environment. Even for $\nu_{\tau}/\nu_{\mu} \, \simeq 1/2$, a mutual 
 comparison of relevant high energy neutrino flux spectra from 
 various neutrino 
 telescopes may be needed to unambiguisly identify the cause of high energy
 neutrino conversions. In case of 
 $\nu_{\mu}\rightarrow \nu_{s}$, a measured ratio of 
 $\nu_{\tau}/\nu_{\mu}\, >\, 10^{-3}$ may provide the corresponding evidence
 thus constituting an observational signature with  relatively large 
 $\delta m^{2}$ ($\delta m^{2}\, \sim \, 10^{-3}$ eV$^{2}$) for high energy
neutrinos originating from AGNs. This follows from the realization that the 
 existing/future high energy neutrino telescopes may measure the ratio 
 $(\nu_{\tau}+\bar{\nu}_{\tau})/(\nu_{\mu}+\bar{\nu}_{\mu})$ \cite{LP}.

Let us note that all these spin (flavour)-precession situations may be 
realized without VEP. In the previous studies \cite{R,W} on 
spin-flip effects for AGN neutrinos, the possible
cause for distortions in the relevant neutrino spectra was attributed to 
either nonzero $\Delta f$ or strong gravitational effects.
However, as we have noticed a  spin (flavour)-precession for  AGN neutrinos 
may develop
even without VEP or/and strong gravitational effects. Thus, the cause 
of change (as compared to no
precession situation) in the ratio of the $\nu_{\tau}$ and 
$\bar{\nu}_{\mu}$ fluxes, 
$\nu_{\tau}/\bar{\nu}_{\mu}$, in existing/future high energy neutrino 
telescopes may not only be attributed to
VEP and/or gravitational effects particularly if an almost energy 
independent permutation for $\bar{\nu}_{\mu}$ and
$\nu_{\tau}$ neutrino spectra takes place for small $\delta m^{2}$ 
with large relevant precession probability. If $\mu$ 
 is of Majorana type then Eq. (4) gives an excess of 
$\nu_{\tau}$ as compared to $\bar{\nu}_{\mu}$ in case of a complete 
 permutation 
as the estimated initial flux of $\nu_{\tau}$ is at least three
orders of magnitude smaller than that of $\bar{\nu}_{\mu}$. On the 
other hand the ratio
 between the $\nu_{e}$ and $\bar{\nu}_{\mu}$ fluxes is at most 
just a factor of 2. Let us note that this situation, if realized 
 empirically will provide an evidence 
 for a violation of total lepton number by two units. 
Conversely, if $\mu $ is of Dirac type then a
deficit in an otherwise large $\bar{\nu}_{\mu}$ flux is expected since
now the precessed  neutrino state is a sterile one. 
 The sterile neutrinos do not interact weakly and therefore are accounted 
 for 
by the disappearance/appearance of the 
 relevant active neutrino ($\bar{\nu}_{\mu}$) flux. 
If the 
 precessions are incomplete (as compared to  complete permutations),
 the resulting high energy neutrino spectra are expected to be rather 
 complicated combinations of the two neutrino species involved in the 
precession.

{\noindent \bf Case 2.} $V_{G}\neq \, 0$ 
(with small $\delta$, that is, $\delta \, \ll \,
2\mu B$). For constant $V_{G}$, we obtain from Eq. (4) the relevant precession 
 probability 
expression by replacing $X$
with $V_{G}$.
 If $V_{G}\, \ll \, 2\mu B$ then using Eq. (1) and Eq. (3), we obtain 
 $\Delta f \, \ll \, 6\cdot 10^{-32}$. We take here  
$|\Delta f|\, {\buildrel < \over {_{\sim}}} \, 10^{-34}$ as our criteria 
and so consequently the corresponding  $P$ reduces to (5). This results 
 in
$P\, >\, 1/2$ with no energy dependence. Thus this case coincides with 
 case 
1a) for small $\Delta f\, ({\buildrel < \over {_{\sim}}} \, 10^{-34})$ 
depending on the given $B_{AGN}$ profile.
	 Consequently, if there is a violation 
of equivalence principle at the level of $10^{-34}$ or less, a 
 spin (flavour)-precession for  neutrinos  may occur in the
vicinity of AGN with small $\delta m^{2}$. Evidently, this value of 
$\Delta f$ is
independent of the gravity mixing angle \cite{G}. For  
$\Delta f\, {\buildrel > \over {_{\sim}}} \, 10^{-34}$, energy 
 dependence in
$P$ results with $P\, {\buildrel < \over {_{\sim}}} \, 1/2$. For large 
$\delta $ ($\delta \, \simeq \, V_{G}$) see next case  and if
$\delta \, \gg \, V_{G}$ then this case reduces to 1c). 
	The upper bound for $\Delta f$ obtained in this case has only a 
linear energy dependence,
 whereas the other necessary requirement [Eq. (6)] does not depend on $E$ 
 for small $\delta$.
 This is in sharp contrast to the situation discussed in next case, 
where both the level crossing as well as the adiabaticity conditions 
 depend
on $E$. Thus, to summarize, we have pointed out in this case that for  
high energy neutrinos originating from AGN, a  spin (flavour)-precession 
 may
develop in the vicinity of AGN if 
$\Delta f \, {\buildrel < \over {_{\sim}}} \, 10^{-34}$ and for a 
reasonable choice of other parameters of the AGN model yielding 
 $\nu_{\tau}/\bar{\nu}_{\mu}\, \gg \, 10^{-3}$.

{\noindent \bf Case 3.} $V_{G}\, \simeq \delta $. This results
in conversion effect in contrast to the previously considered two
 cases [which are spin (flavour)-precession effects]. 

	Two conditions are essential for an adiabatic conversion:
 i) level crossing and ii) adiabaticity. The level crossing is 
 obtained by equating the diagonal elements of the effective 
 Haliltonain in Eq. (2), that is, $V_{G}\, =\, \delta$.  
The level crossing for $E\, \sim \, $1 PeV implies
\begin{equation}
 10^{-39}\left(\frac{\delta m^{2}}{10^{-10}eV^{2}}\right)\, 
 {\buildrel < \over {_{\sim}}} \Delta f \, 
 {\buildrel < \over {_{\sim}}} \,  
 10^{-29}\left(\frac{\delta m^{2}}{1 eV^{2}}\right).
\end{equation}
Note that relative sign between $\delta $ and $V_{G}$ is important for level 
crossing. From the level crossing it follows that 
$\Delta f\, \propto \, E^{-2}$, i.e.,
an inverse quadratic $E$ dependence on $\Delta f$. 
 This level crossing is induced by an interplay of magnetic field and VEP.
 However, level crossing alone is not a sufficient condition for a complete
 conversion. As stated earlier, adiabaticity is the other necessary 
 condition that determines the extent of conversion. If there is only level
 crossing and no adiabaticity at the level crossing then there is no 
conversion of antimuon neutrinos into tau (sterile) neutrinos.   
The relevant adiabaticity condition  may be written as 
\begin{equation}
 B_{\mbox{ad}}\,
 {\buildrel > \over {_{\sim}}} \, 3\cdot 10^{2}\, 
 \mbox{G}\left(\frac{\Delta
 f}{10^{-29}}\right)^{\frac{1}{2}}\left(\frac{10R_{S}}{r}\right).
\end{equation}
We note that $B_{\mbox{ad}}\, {\buildrel < \over {_{\sim}}}
 \, B_{AGN}$ for 10$\, <\, x\, <\, $100 with $\mu \, \sim \, 
 10^{-12}\mu_{B}$. Thus, the adiabatic conversion may occur 
giving rise to energy
dependent distortions with corresponding conversion probability greater 
than 1/2. For large $\delta $
whereas a  spin (flavour)-precession is suppressed [see case
1b) and 1c)], an adiabatic conversion may result with $P\, >\, 1/2$ for 
large $\Delta f$ (see Fig. 1). Let us emphasize that this adiabatic level 
 crossing is induced by the change in the gravitational potential rather
 than the change in effective matter density. Thus, observationally, we
may obtain here $\nu_{\tau}/\bar{\nu}_{\mu} {\buildrel > \over {_{\sim}}} \,
 1$ due to an adiabatic conversion induced by an interplay of $V_{G}$ 
 and $B_{AGN}$ in the vicinity of the AGN. For $\delta \, \ll \,
V_{G}$ this case reduces to case 2 whereas for $\delta \, \gg \, V_{G}$, 
 we obtain case 1c).

\begin{figure}[htb]
\vspace{7.2cm}
\includegraphics{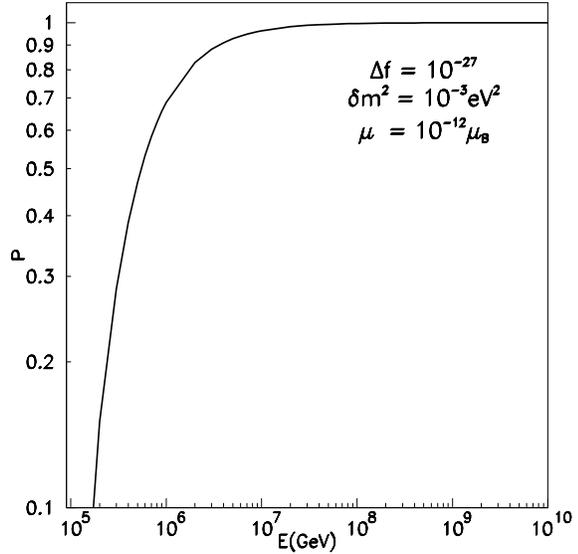}
\caption{$P(\bar{\nu}_{\mu}\, \rightarrow \, \nu_{\tau})$ Vs $E$ 
 including the effect of possible nonadiabaticity.
\label{fig:tau}}
\end{figure}

	It follows from the discussion in case 3  that a 
{\em nonzero} $\Delta f$ is needed to induce an adiabatic level crossing
 with $P\, >\, 1/2$. It is in contrast to cases 1 and 2 where a 
 spin-flip may occur through spin (flavour)-precession without $\Delta f$ 
 with $P\, >\, 1/2$.   

\section{RESULTS AND DISCUSSION}

	1. The initial fluxes of the high energy neutrinos 
originating from AGN are estimated
to have the following ratios: $\nu_{e}/\nu_{\mu}\, \simeq 1/2$, $
\nu_{\tau}/\bar{\nu}_{\mu} \, {\buildrel < \over {_{\sim}}} 10^{-3}$. 
 Thus, if an enhanced $\nu_{\tau}/\bar{
\nu}_{\mu}$ ratio (as compared to no precession/conversion situation) is 
observed correlated to the direction
of source for high energy 
neutrinos, then it may be either an evidence for a  spin-flip through 
spin (flavour)-precession alone or through 
 a resonant conversion in the vicinity of AGN due to
an interplay of VEP and $B_{AGN}$ depending on the finer 
 details of the relevant high energy AGN neutrino spectra.
 The spin (flavour)-precession and/or conversion effects  
discussed in this study  may be distinguished by observing
the energy dependence of the high energy neutrino flux profiles. 
 A mutual comparison of the relevant (that is, $\bar{\nu}_{\mu}$ and 
$\nu_{\tau}$) high energy neutrino spectra may in 
 principle isolate the mechanism of neutrino 
conversions in the vicinity of AGN.

	2. For small 
$\delta m^{2}$ ($\delta m^{2}\, <\, 5\cdot 10^{-6}$ eV$^{2}$) 
a  spin (flavour)-precession may result in an energy
independent permutation of the relevant neutrino spectra with the 
corresponding spin (flavour)-precession
probability greater than 1/2. This spin (flavour)-precession may 
 occur for small 
$\Delta f$ ($\Delta f\, {\buildrel < \over {_{\sim}}} \, 10^{-34}$). 
 The spin-flip may occur through resonant conversions induced by the 
VEP and $B_{AGN}$ as
well. Assuming that the information on $\Delta f$ may be obtained from
various terrestial/extraterrestial experiments, a mutual comparison 
 between the survived and transformed high
energy AGN neutrinos may enable one to distinguish the mechanism of 
conversion. If for small
$\delta m^{2}$ ($\delta m^{2}\, <\,
5\cdot 10^{-6}$ eV$^{2}$), an energy dependent permutation between 
$\bar{\nu}_{\mu} \, (\bar{\nu}_{e})$ and
$\nu_{\tau}$ are
obtained empirically with corresponding $P\, >\, 1/2$ then this 
 situation may 
 be an evidence for a conversion effect
due to an interplay of VEP and  $B_{AGN}$. 

	For large 
$\delta m^{2}$ ($\delta m^{2}\, >\, 5\cdot 10^{-6}$ eV$^{2}$), 
if  energy dependent
distortions (and a change in $\nu_{\tau}/\bar{\nu}_{\mu}$) is observed 
 with the corresponding conversion
probability greater than 1/2 then the cause may be a relatively large 
$\Delta f$ ($\Delta f\, >\, 10^{-34}$). Thus, 
with the improved information on $\Delta f$  the cause of the
conversion effect may be isolated. Further, as the energy span in 
 the relevant high energy AGN neutrino spectra is several  orders of
magnitude, therefore, energy dependent spin (flavour)-precession/conversion 
 probabilities may result in distortions in
some part(s) of the spectra for relevant neutrino species and may thus be 
identifiable in existing/future high energy neutrino telescopes.

\section*{ACKNOWLEDGEMENTS}

The author thanks
 R. A. V\' azquez for useful discussion and
 Agencia Espa\~nola de Cooperaci\'on Internacional (AECI), Xunta de 
 Galicia
(XUGA-20604A96) and CICYT (AEN96-1773) for financial support.

\end{document}